# IMPLEMENTATION OF A REAL TIME PASSENGER INFORMATION SYSTEM

**Ganesh K**

Net Logic Semiconductors Pvt. Ltd., Bangalore, India   E-mail: ganesh.iisc@gmail.com

**Thrivikraman M**

Rambus Technology Inc., Bangalore, India E-mail: vikram.iisc@gmail.com

**Joy Kuri***

Center for Electronics Design and Technology,
Indian Institute of Science, Bangalore, India   E-mail: kuri@cedt.iisc.ernet.in
***Corresponding Author**

**Haresh Dagale**

Center for Electronics Design and Technology,
Indian Institute of Science, Bangalore, India   E-mail: haresh@cedt.iisc.ernet.in

**Sudhakar G**

Amber Root Systems Pvt. Ltd., Bangalore, India. E-mail: sudhakar@amberroot.com

**Sugata Sanyal**

School of Technology and Computer Science,
Tata Institute of Fundamental Research, Mumbai, India. Email: sanyals@gmail.com

## ABSTRACT

Intelligent Transportation Systems (ITS) are gaining recognition in developing countries like India. This paper describes the various components of our prototype implementation of a Real-time Passenger Information System (RTPIS) for a public transport system like a fleet of buses. Vehicle-mounted units, bus station units and a server located at the transport company premises comprise the system. The vehicle unit reports the current position of the vehicle to a central server periodically via General Packet Radio Service (GPRS). An Estimated Time of Arrival (ETA) algorithm running on the server predicts the arrival times of buses at their stops based on real-time observations of the buses' current Global Positioning System (GPS) coordinates. This information is displayed and announced to passengers at stops using station units, which periodically fetch the required ETA from the server via GPRS. Novel features of our prototype include: (a) a route-creator utility which automatically creates new routes from scratch when a bus is driven along the new route, and (b) voice tagging of the stops and points of interest along any route. Besides, the prototype provides: (i) web-based applications for passengers, providing useful information like a snapshot of present bus locations on the streets, and (ii) web-based analysis tools for the transport authority, providing information useful for fleet management, like number of trips undertaken by a specific bus. The prototype has been demonstrated in a campus environment, with four-wheelers and two-wheelers emulating buses. The automatic real-time passenger information system has the potential of making the public transport system an attractive alternative for city-dwellers, thereby contributing to fewer private vehicles on the road, leading to lower congestion levels and less pollution.

# 1. INTRODUCTION

With the advent of GPS and the ubiquitous cellular network, real time vehicle tracking for better transport management has become possible. These technologies can be applied to public transport systems, especially buses, which are not able to adhere to predefined timetables due to reasons like traffic jams, breakdowns etc. The increased waiting time and the uncertainty in bus arrival make public transport system unattractive for passengers. A Real-Time Passenger Information System (RTPIS) uses a variety of technologies to track the locations of buses in real time and uses this information to generate predictions of bus arrivals at stops along the route. When this information is disseminated to passengers by wired or wireless media, they can spend their time efficiently and reach the bus stop just before the bus arrives, or take alternate means of transport if the bus is delayed. They can even plan their journeys long before they actually undertake them. This will make the public transport system competitive and passenger- friendly. The use of private vehicles is reduced when more people use public transit vehicles, which in turn reduces traffic and pollution.

Reference [1] describes some of the existing RTPISs in different parts of the world -- UK, USA, Ireland, Taiwan and Italy. It also compares them in terms of the vehicle location technology, ETA prediction and mode of information dissemination. Reference [2] describes details about a European RTPIS project, INFOPOLIS 2. It presents a survey about different passenger information systems present in European countries. Reference [3] describes a project study in implementing passenger information system in the city of Los Angeles. Reference [4] describes the RTPIS of Helsinki, Finland. Reference [5] presents a commercially available passenger information system that has features like information delivery via SMS (Short Messaging Service), webpages, on-board and at-stop displays, location-based advertisements, alerts about schedule changes and a journey planner. Reference [6] is another commercial system with only onboard and at-stop displays for next stop and waiting time display. References [7]-[13] describe arrival time prediction algorithms. Reference [14] describes the preferred mode of information for passengers in buses, at stops and passengers who are planning to travel. Reference [15] and [16] describe future work plan, related to security issues and also about how to apply this methodology for other possible fleet management.

The application scenario of RTPIS is shown in Figure 1.

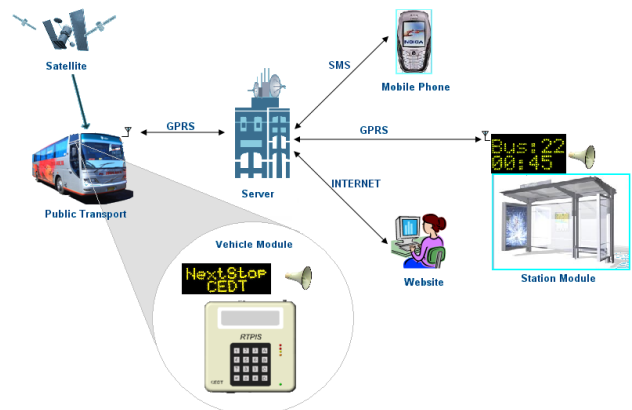

**Figure 1  RTPIS Scenario**

RTPIS provides travel information to passengers and tourists enabling them to make informed decisions about modes, routes and departure times. The RTPIS framework can be broadly divided into two contexts: Pre-trip context and On-trip context. The former provides information like timings, fares and routes well before the commencement of travel, through the Internet or the Short Messaging Service (SMS). The On-trip context provides information like location and places of interest (POI) while on the move. This is achieved using on-board and at-stop terminals (displays and audio announcement units).

Novel features of our prototype include:
- **Route creation utility:** Information about bus routes can be built up from scratch using this. The new route is created automatically when the bus is driven along the new route. The number of nodes used to describe a route is determined adaptively, depending on whether the route features many short linear segments, or few large linear segments. To the best of our knowledge, the route creation feature of our system is *not found* even in commercial implementations.

- **Voice tagging:** Our system enables the driver to tag a bus stop with a name, by recording its name during the process of route creation. This voice tag will be replayed when the bus approaches the bus stop next time.

The paper starts with an overall description of the system implementation (Section 2). The route creation and voice tagging features are covered next (Section 3). ETA is the main travel information provided to passengers. Using the current known position of the buses and their routes, the ETA predictor (Section 4) calculates the ETA at every bus stop. A host of different server utilities have been developed which includes, a relatively new feature of ETA retrieval by SMS (Section 5). In Section 6, we provide estimates of the server load and fleet size that can be managed by the system. Section 7 covers future work plan.

## 2. SYSTEM DESCRIPTION

The main parts of RTPIS are vehicle units in buses, station units at bus stops and a central data processing server. These parts are briefly described in the subsequent sections.

### A. Vehicle unit
The main functions of the vehicle unit are as follows.

#### Pre-trip context
- To download names and coordinates of stops and points of interest from the server

#### On-trip context
- To compute current location, direction and speed of the bus.
- To transmit the computed information to the central server using GPRS.
- To display "next stop/point of interest" information on the vehicle, and play out corresponding announcements
- To provide a "user interface" consisting of a display and keypad for the driver, enabling bus management actions, like route number changes and breakdown indications.

The block diagram and photograph of the vehicle unit are shown in Figure 2 and Figure 3 respectively.

The hardware platform for this unit is described in Appendix A.

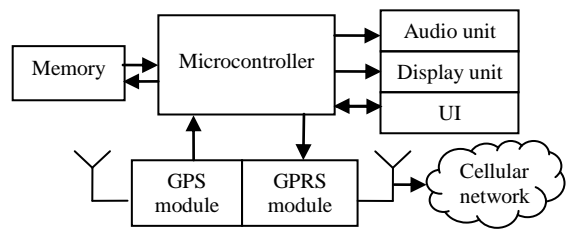

**Figure 2 Vehicle unit Block Diagram**

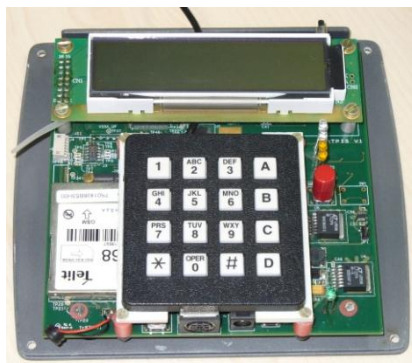

**Figure 3 Photograph of vehicle unit**

The vehicle unit operates as follows – the GPS receiver in this unit computes the current location of the vehicle. The latitude, longitude and speed of the bus are transmitted periodically to a central server using GPRS. The vehicle unit initially downloads the names and coordinates of stops and POIs on the current route from the server. This is used to display and announce the name of stop or POI when it is approached, for the benefit of passengers inside the bus. The configuration parameters and audio files for announcement are stored in memory. The microcontroller sequences the series of operations --- periodic transmission of GPS coordinates, audio and display operations. The UI (keypad and display) is meant for the bus driver, and is used to change the route number of the bus, indicate breakdown and trace a new route.

The Firmware for the vehicle unit consists of a Real Time Operating System (RTOS), application tasks that run on the microcontroller and Python scripts that run on the GPS/GPRS module which is used as a coprocessor. RTOS has been used in the vehicle unit for modular firmware development. Functions related to positioning and communications with the server are handled by Python scripts running on the GPS/GPRS module, and microcontroller application tasks handle the remaining functions of the vehicle unit. This task partitioning across microcontroller and the module reduces load on the former. This is significant because when audio files are being played, the microcontroller will not be able to process the position/ETA updates efficiently.

*B. Server*

The server is at the center of our RTPIS. The functions of the server are listed below:

- To maintain a database of all the routes, the buses that ply on a route, the stops along each route etc. (Table 1)
- To continuously receive location and speed from the vehicle units of all the buses
- To calculate the ETA of all the buses at their next and subsequent bus stops
- To reply to SMS-based queries requesting ETA at specific stops from users; a GSM modem connected to the server transfers these queries to the server which processes them and sends the reply message
- To host Internet web pages, which allow users to track buses in real time, see the route map of any route, and get the ETA for any route-stop pair and plan trips from any source to any destination stop, at any time

*1) Server database*

The server maintains a database of information pertaining to the buses, routes and stops in the form of tables (Table.1). The server database [12] can be organized in many ways, to reduce memory requirement, improve access speed, or reduce the number of queries. To improve the query speed, the tables related to buses are partitioned into static and dynamic ones. The Bus table stores static data while the bus position and log tables store dynamic data. The relation between the unique bus id, bus type (ordinary/luxury/…) and route number is stored in the Bus table. The position updates from the bus are stored in the Bus Position and the Bus Position Log tables. The "direction" field indicates the direction in which the bus is headed (Terminus A to Terminus B, or reverse). The direction is calculated in the vehicle unit by comparing time-separated position values with route details. The average speed is the weighted average of the current speed and the previous average speed. The status of the bus changes to invalid, when its driver signals a breakdown. This helps the transport company to take suitable actions. The bus is excluded from ETA calculations based on this field. The Bus Position Log table stores a copy of the position update sent by buses. This log can be backed up and used for future analysis (Analysis tools, section 5).

TABLE 1: DATABASE TABLES

| Table name | Contents |
|---|---|
| Bus | Bus ID, type and route |
| Bus Position | Bus ID, current coordinates, speed, average speed, direction, current link, link entry time, estimated end time, status |
| Bus Position Log | Log of changes made to bus position table |
| Node | Node ID, coordinates, name |
| Link | Link ID, node pair, travel time |
| Route | Route number, sequence of links |
| Stop | Stop name, node ID, route number, ETA |

To store the route information, a route is modelled as a set of nodes, which can be stops, POIs or critical bends. Adjacent nodes form links. Thus, a route is characterised by an ordered set of links as in Figure 4. The current link position of each bus and the estimated time to reach the link end are calculated and stored in the Bus Position table upon reception of an update from the vehicle unit.

Details about the links and nodes are stored in the Link and Node tables, respectively. The link travel time is dynamically updated when a bus reaches the end of the link. Each stop has details of buses that pass through them along with their estimated time of arrival at the stop.

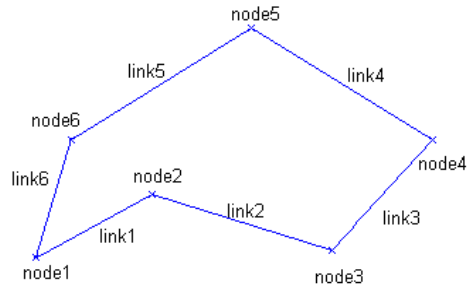

**Figure 4 A typical route identified by links**

### C. Station unit

The station units are installed at all major bus stops. The station unit functions are listed below.

- To fetch ETA for all routes through the stop
- To announce and display the fetched ETA

The architecture, as shown in Figure 5, is similar to that of the vehicle unit, except for the absence of the GPS receiver and the UI. Reuse of the same PCB design reduces the manufacturing cost. The station unit operates as follows. The GPRS module periodically fetches ETA information for all routes through the stop, from the server via GPRS and the microcontroller sequences this information to the audio and display units. Since the station unit is similar to the vehicle unit, the PCB designed for the vehicle unit, shown in Figure 11, is reused. This is done by using a GPRS-only module in place of GPS/GPRS module and not mounting the Liquid Crystal Display (LCD) and keypad.

As for the vehicle unit, the firmware for the station unit is split into microcontroller application tasks for control and sequencing, and Python scripts for communication with server.

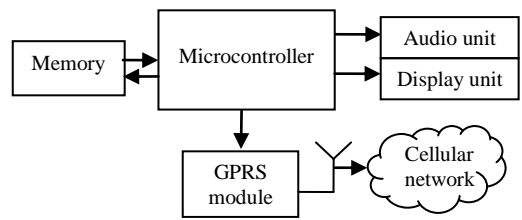

**Figure 5 Station unit Block Diagram**

### 3. ROUTE CREATION

A novel method has been developed to automate the process of creating new routes and populating the database, with little human intervention. To create a new route, the vehicle unit is taken along the required route. The unit keeps logging the position coordinates along the route in the local memory at a fast rate. Whenever a stop needs to be indicated, the driver can attach a voice tag to the position information. The tag serves the purpose of attaching a name to a coordinate later. A server utility takes the data collected by the vehicle unit and creates new tables in the database for the created route. This program applies a piece wise linear (PWL) approximation algorithm to the raw set of nodes, and optimises the number of nodes and links by retaining only the critical nodes. The algorithm has the built-in feature that nodes are introduced whenever required, according to the threshold set by the user (the transport authority). If the route involves many turns (zigzagging route), then more nodes are introduced; if the route involves long stretches of straight lines, then fewer nodes are introduced. A smaller error threshold results in accurate representation of the route.

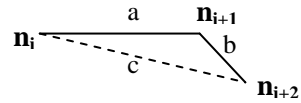

**Figure 6 A node tuple used in the route creation algorithm.
a, b and c denote the distances between the indicated nodes**

The algorithm considers three nodes ($n_i$, $n_{i+1}$, $n_{i+2}$) at a time and checks how collinear they are. If the linearity error (defined as a+b-c in Figure 6) is less than a threshold (25m in the implementation), the second node is skipped and a new node tuple ($n_i$, $n_{i+2}$, $n_{i+3}$) is considered. If the error crosses the threshold, the second node in the tuple is added to the list of nodes, and the algorithm repeats starting with the second node in the tuple. However, all stops are included irrespective of the linearity constraint. The pseudocode for the route creation algorithm is given in Appendix B.

After optimising the set of nodes (Figure 7), node and link information must be entered in the database. For this, the new nodes are compared with the existing nodes in the database. If there is a match (their distance is within a threshold, 25m in the implementation), the existing node is reused. If no match is found, an entry for the node that does not exist in the database is added to the Nodes table. Similarly, existing links are checked for matches. If a link is already in the database, it is reused; else, a new link is added to the Link table.

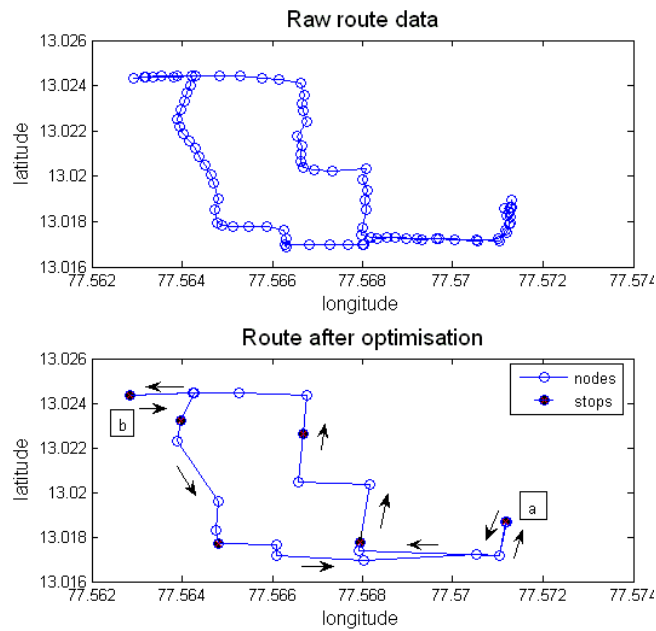

**Figure 7 Route creation**

### 4. ETA PREDICTION

Arrival time prediction forms the core of any RTPIS system. The algorithm can be very simple, involving only a bus schedule table [7], zone based [8] or could be very complicated, involving Artificial Neural Networks [9], space-time correlation [10] and time series modeling [11]. Bus schedule table and past location data can be used to predict arrival time, as in [13].

Our system provides a platform for executing *any* ETA algorithm, though we have implemented our own simple one that adapts to changing traffic conditions. Our algorithm works by recording the time it takes to traverse each link. Predictions are based on the present and past observations of a bus passing through each link. The past observations get lesser weight as time progresses; this reflects current traffic conditions better. The pseudocodes for these algorithms are given in Appendix B.

The predicted ETA at bus stops is bounded by an upper limit of one round trip time of the route, though the ETA can be predicted infinitely into the future by simply adding integral number of round trip times to the smallest ETA.

The ETA algorithm has two parts:

1. Link updater, which estimates the travel time for each link
2. ETA calculator, which calculates the ETA for every bus stop.

### D. Link Updater

Link updater calculates the link travel times required by the ETA calculator. Whenever a bus position update is received from the vehicle unit, the link updater calculates the travel times for all links traversed by the bus from the previous known position. The weighted average of the previous value and the actual travel time obtained for the current bus is stored as the link travel time in the Link table. To compute the average velocity, the weights are 90% for the previous average velocity and 10% for the current velocity. To compute average link travel time, the weights are 70% for the previous average and 30% for the current value. For an update rate of two per minute used in our trial runs, these weights give a good approximation of the average values, as well as track the recent trends. The link travel time is also common to all routes containing the link, so as to get the latest time estimate. This is the reason for sharing links between routes during route creation.

Link updater locates the bus position along the current route of the bus. The link updater then calculates the time required to reach the end of the current link and updates the estimated end time information in the Bus Position table. If the bus enters a new link, the entry time for the new link is stored in the Bus Position table against the bus and the travel time for all the crossed links is calculated. This time is also the exit time for the previous link. The time difference between the exit time and the previously recorded link entry time gives the link travel time for the crossed links. The travel times for links are a function of their lengths. Thus, when more than one link is traversed between updates, the individual link travel times are computed as *fractions* of the total travel time, with the fraction for link $i$ being the ratio of the length of the $i^{th}$ link to the sum of lengths of traversed links. This makes sure that among the traversed links, shorter links have smaller travel times and longer links have larger travel times. The computed link travel times are averaged with their previous values and the Link table is updated.

Suppose that the last position update received from a bus corresponds to position $B_I$ (Figure 8), and the position update received just now corresponds to position $B_F$. We discuss the actions of the link updater algorithm for this example. The route section consists of the links $N_1$-$N_2$, $N_2$-$N_3$, $N_3$-$N_4$, $N_4$-$N_5$, where $N_i$ are the nodes comprising the route. In the first pass of the link updater algorithm, the latest bus position is mapped onto the route by checking every link in the route, starting from the current link. In this case, the bus is found to be in the link $N_4$-$N_5$. Since it is a new link, the entry time into that link (for this bus) is set to the current time. In the second pass of the algorithm, travel time for the crossed links --- $N_1$-$N_2$, $N_2$-$N_3$, $N_3$-$N_4$ --- are updated by partitioning the total travel time among individual link travel times, based on their lengths. Hence, link $N_1$-$N_2$ gets a larger travel time than link $N_2$-$N_3$. For the final link, the time to reach the end of the link (node $N_5$) is found by dividing the Euclidean distance between the bus position and $N_5$, by the average velocity of the bus. Database updates for bus position, current link, link travel times, estimated end time and average velocity happen at every iteration of the algorithm.

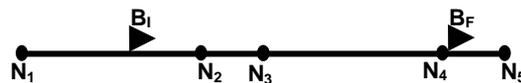

**Figure 8 Link travel time calculation scenario**

### E. ETA calculator

This program takes the current bus position, link travel times and estimated time to link-end to predict the ETA for all bus stops. ETA at a stop is the time taken for the nearest bus to reach the bus stop. It is calculated as the sum of travel times of the links, starting from the current bus position, up to the given bus stop.

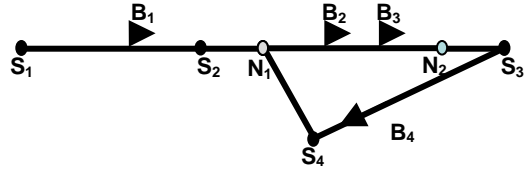

**Figure 9 Scenario for ETA calculation**

An example scenario for ETA calculation for a route is shown in Figure 9. The *circular* route consists of links $S_1$-$S_2$, $S_2$-$N_1$, $N_1$-$N_2$, $N_2$-$S_3$, $S_3$-$S_4$, $S_4$-$N_1$, $N_1$-$S_2$, $S_2$-$S_1$ where, $N_i$ are intermediate nodes and $S_i$ are stops. ETA calculation starts by checking for buses in links of the route. In the example, a bus is present in the first link $S_1$-$S_2$ and the end node of the link is a stop. Hence, the ETA at stop $S_2$ is just the estimated time to link-end for the bus $B_1$. Since no bus is present in link $S_2$-$N_1$, ETA at $N_1$ is the sum of ETA at $S_2$ and the link travel time of $S_2$-$N_1$. In the link $N_1$-$N_2$ there are 2 buses and bus $B_3$ is closer to $N_2$ than $B_2$, and hence ETA at $N_2$ is estimated time to link-end of bus $B_3$. ETA at other nodes is calculated in a similar manner and they are listed in Table. 2.

**TABLE 2: EXAMPLE ETA VALUES**

| Stop | ETA |
| --- | --- |
| $S_1$* | ETA($S_2$, return) + ltt($S_2$-$S_1$)** |
| $S_2$ | Estimated-end-time($B_1$) |
| $S_3$ | Estimated-end-time($B_3$) + ltt($N_2$-$S_3$) |
| $S_4$* | Estimated-end-time($B_4$) |
| $S_2$,return* | ETA($S_4$) + ltt($S_4$-$N_1$) + ltt($N_1$-$S_2$) |

*eta is for returning bus $B_4$
**ltt – link travel time

## 5. SERVER UTILITIES

Internet based utilities have been developed to help online users get travel information. This has been designed along the lines of the Helsinki City Transport website [4].

### F. Website
Pre-trip information, like routes connecting different places, route map, current position of buses and bus arrival times are provided to users through the RTPIS website. Additional web pages have been created for the system administrators to create new routes, perform bus-wise and route-wise analysis and manage the RTPIS database.

### G. Trip Planner
Trip planner allows passengers to plan their journeys beforehand. The web page allows passengers to enter the source and destination stops and the start time for the trip. For enquiries regarding potential trips, the trip planner finds the routes that are common to the source and destination stops. It then chooses the predicted ETA which is closest in time to the time of interest, from the list of ETAs separated by integral multiples of round trip times.

### H. Bus Arrivals
Through this web page, users can request ETA at a bus stop, for any or every route passing through it. This page queries the database for ETA and displays it to the user.

### I. Route Map
Through this web page, users can request to view the map of the selected route. The displayed map shows the forward and reverse paths and stops along the route. The Google maps Application Programming Interface (API) has been used for this purpose as in Figure 10.

### J. Tracking Buses
Through this web page, users can view the present position of all the buses for the selected route, on the route map. This is done by getting the position of all the buses of a route from the database and then plotting it on the route map.

*K. Analysis Tools*

Using the bus position log table, information regarding the usage pattern of buses and routes can be obtained. This page enables the administrator to find the number of trips a specific bus has undertaken, the total distance travelled and adherence to schedule by drivers. Similarly, information for a particular route, like buses that used that route and number of trips made in that route can be obtained. This helps to develop the infrastructure of critical routes.

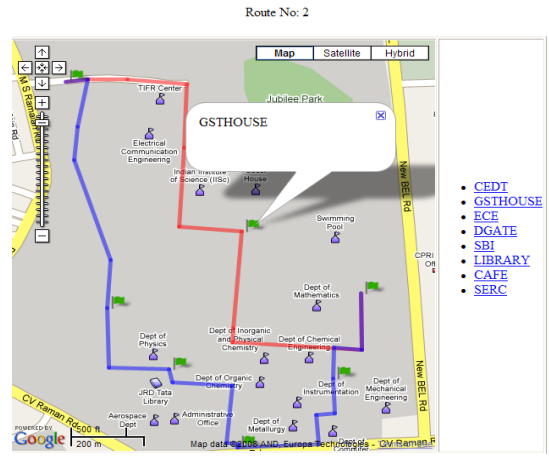

**Figure 10 Route map overlaid on Google maps**

*L. Database Manager*

This page provides unrestricted access to the RTPIS database and hence it is password-protected. Using this page, the administrators can create, edit and delete any table in the database.

*M. SMS Handler*

SMS requests from users are handled by this application running at the server. In the SMS request, users should mention the name of the stop for which they want the ETA and optionally, give their desired route number. After querying the database for the ETA, the results are sent back to the sender via SMS.

## 6. ESTIMATING SERVER LOAD AND FLEET SIZE

The RTPIS system was tested in the Indian Institute of Science campus. Route used for the test is shown in Figure 10. The source stop is CEDT and the destination stop is DGATE. The onward path consisting of the stops CEDT, Guest House, ECE, and DGATE is shown in red, and the return path consisting of the stops DGATE, SBI, Library, Café, and SERC is shown in blue. Initially, when the vehicle unit is powered up at CEDT, it downloads the route information from the server and waits for a GPS position fix. After getting the position fix, it identifies itself to be at CEDT. When the vehicle moves away from CEDT, the direction is automatically inferred as "onward." This helps the vehicle unit to match the current vehicle coordinates with the stops of the route, so as to identify the next stop. The vehicle unit announces the name of the next stop and displays it, about 50m before the stop. When the destination is reached, the direction is automatically reversed to "return path," and the process repeats. Meanwhile, the station unit placed at CEDT announces and displays the ETA at CEDT, and the present position of the vehicle is shown, live, on the tracking webpage of the server. The campus wide test is the source of the experimental data.

RTPIS capacity is limited by the number of queries to the centralised MySQL server. Link updater and ETA calculator are very processing- intensive tasks. Link updater has to query the Bus, Bus Position, Route, Link and Node tables to find link travel times, current link, estimated end time, etc. The ETA calculator has to query the Bus, Route, Link, Node and Stop tables to find the link travel time and the estimated end time, and then update the ETA. This section studies the complexities of algorithms in terms of the number of queries to the server, and arrives at an estimated fleet size based on reasonable assumptions.

*N. Server load estimation*
<u>Notations</u>
B, L, R, S – number of buses, links, routes and stops, respectively, in the database

We use the big 'O' notation to quantify the complexity in terms of the number of queries generated. For instance, O(x) signifies that the number of queries is bounded by a linear function of x.

*1) Load from Link Updater*

For every position update, the link updater tries to locate the bus in one of the links of its route. To do this, it retrieves the links which form the route and has a worst-case complexity of O(L). But most of the buses are likely to be in the next few links from the current link, resulting in a practical value of O(1). Since link updater works whenever a bus position update occurs, resulting in a total complexity of O(B) in practice.

*2) Load from ETA calculator*

For every execution, the ETA calculator queries the database for all the links for every route and has a complexity of O (L).

*3) Load from SMS handler*

For every SMS requesting ETA at a stop, the SMS handler queries the corresponding stop table for the required route. If no route is mentioned, it retrieves the ETA for all routes corresponding to the stop in one query. Hence, the complexity is O(1) for each SMS handled. The query complexity of the website program is similar to SMS handler.

*4) Load from Station unit*

Every station unit retrieves all the ETA for a stop in one query, resulting in a complexity of O(1). For all the station units the total load is O(S).

*O. Fleet size estimation*

Fleet size can be estimated by relating the load on the server and the capacity of the server. To arrive at the actual load, the big 'O' notation is replaced by scaled values of the complexity, i.e. O(x) by kx. The mathematical relation and the fleet size estimate are described below.

*Assumptions*

Server capacity, $r_{SERVER}$ = 30000 queries/min
Position update rate for a bus, $r_B$ = 2 /min
ETA calculation rate, $r_{ETA}$ = 1 /min
No. of SMS queries, $r_{SMS}$ = 20 /min
No. of queries from website, $r_W$ = 50 /min
No. of queries from a station unit, $r_S$ = 1 /min

**Total no. of queries generated for all buses, routes, stations, user requests etc...:**

Link updater = $\alpha r_B B$ /min
ETA calculator = $\beta r_{ETA} L$ /min
SMS handler and website = $\gamma (r_{SMS} + r_W)$ /min
Station units = $\delta r_S S$ /min

The server capacity number was arrived at experimentally by giving continuous updates covering one route to the server and finding the execution time. For faster servers, this number can be much higher. α is the observed value, from the execution of "current link updater" program. β, γ and δ are obtained mathematically based on the algorithmic implementation.

Number of queries generated per minute by the whole system should be less than the total number of queries the server can handle.

$\alpha r_B B + \beta r_{ETA} L + \gamma(r_{SMS} + r_W) + \delta r_S S < r_{SERVER}$

For a typical case of 100 links per route, 10 buses per route, and 2000 stops in all, and for the scale factors of α≈10, β≈1, γ≈6, δ≈1, we get number of routes, R ≈ 91 and number of buses, B ≈ 910.

## 7. FUTURE WORK

As this system uses a combination of processing elements: PCs, Mobile Phones etc., there is a possibility of the overall system malfunction due to a particular type of attack, it is termed as Denial of Service (DoS) attack by malicious agents who might try to disrupt the function of the system. Reference [15] describes a Distributed Security Scheme for Ad Hoc Networks and it also proposes a proactive scheme to prevent this kind of attack. Similar methodology will be studied to make this Real Time Passenger Information System more robust.

The proposed system is also quite universal in nature and it is possible to extend the methodology for other type of fleet movement where security is of paramount importance. Reference [16] proposes a novel data hiding technique, based on Steganographic mechanism. Here, the advantage lies in the fact that computationally costly encryption-decryption mechanism is avoided, thus making it suitable for a heterogeneous combination of processing elements, which are being used in our present system. Here, many processing elements e.g. Mobile phone etc. lacks the processing power and battery power, which is required for traditional encryption-decryption system.

## 8. SUMMARY

We have developed a RTPIS that tracks the current location of all the buses and estimates their arrival time at different stops in their respective routes. Estimates are updated every time the bus sends an update. It distributes this information to passengers using display terminals at bus stops, web based GUI and SMS. This system serves the needs of passengers, vehicle drivers and administrators of the transport system. The system can be deployed in any city with approximately 1000 buses, 100 routes and 2000 stops. New routes can be added without the need for detailed maps, using a user friendly GUI and data from vehicle drivers. Starting from scratch, all routes can be added to the database, in a few days without any major data entry work at the server. In addition, real time traffic data from link travel time statistics can also aid city traffic information systems.

## APPENDIX

*P. PCB block schematic*

The block schematic of the four-layer PCB designed for the vehicle unit is shown in Figure 11. Vehicle and station units have the common functionalities of announcements, display and communication. The vehicle unit has the additional requirement of calculating the current position. The components of the PCB have been chosen to perform the above said functionalities.

The microcontroller (µC) is connected to the combo GPS/GPRS module through the Universal Asynchronous Receiver Transmitter (UART) interface. AT commands are issued across this interface, to get GPS coordinates and perform GPRS data transfer. The Secure Digital (SD) card adapter interfaces an SD card to the µC, via the Serial Peripheral Interface (SPI). The codec is interfaced to the µC through the Inter Integrated Circuit ($I^2C$) interface for configuring the codec registers and SPI for transferring audio data. The SD card and codec are jointly used to play and record audio files. LCD and keypad are interfaced through General Purpose Input Outputs (GPIO) and provide a user interface to the driver. The power section generates the necessary supply voltages for all the components from the 12V input. A DIN connector carries data for the external LED display and speaker. The JTAG interface is to program the µC.

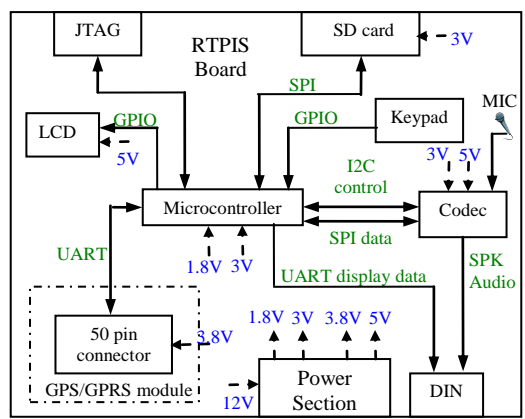

**Figure 11 RTPIS PCB block schematic Algorithms**

**General notation -**
i, j, n, p : any node
dist(i,j) : Euclidean distance between i and j
error($n_0,n_1,n_2$) : dist($n_0,n_1$)+dist($n_1,n_2$)-dist($n_0,n_2$)
l         : any link
start(l) : first node of link l
end(l) : second node of link l
$d_l$     : length of link l
$TT_l$   : time to traverse link l

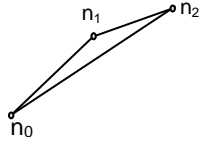

1) **PWL Algorithm**

a) *Notations*
   n [ ] - list of nodes from node file
   ñ [ ] - list of optimised nodes

b) *Initialisations*
   N ← length (n)
   ñ[0], $n_0$ ← n[0]
   $n_1$ ← n[1]
   j  ← 0, link counter
   l  ← [], list of links
   err ← 0, accumulated error

c) *Algorithm*
   for i ← 2 to N-1
      $n_2$ ← n[i]
      err = err + error($n_0,n_1,n_2$)
      if $n_2$ is stop/POI or err > threshold
         err = 0
      ñ [ j+1] ← $n_1$
         l[j] ← link $n_0$- $n_1$
         j ← j + 1
         $n_0$ ← $n_1$
      $n_1$ ← $n_2$

2) **Link Updater**

a) *Notations*
   clnk ← last known link of the bus
   p ← position of bus, sent by the vehicle unit
   lnk  ← list of links that need to be updated
   LST ← link entry time for clnk
   Δt  ← current time – LST
   $d_T$ ← total length of completed links
   threshold ← constant, 60m in implementation

b) *Algorithm*
   //pass1- find the links to be updated
   $d_T$ ←  0
   for l ∈ links in route starting from clnk
      if error(start(l), p, end(l)) < threshold
         new_clnk ← l
         $d_T$ ← $d_T$ + $d_l$
         if new_clnk != clnk
       LST ← current time
   //pass2 - update the links
   avg_vel = avg_vel * a + cur_vel * (1-a)
   lnk ← set of links along the route from clnk to new_clnk

```
    for l ∈ lnk
        if  l != new_clnk
            TT_l ← b*TT_l + (1-b)* Δt* d_l / d_T
        else
            clnk ← new_clnk
            est_end_time ← dist(p, node2(new_lnk)) / avg_vel
```
where, a and b are the averaging factors, implementation values are a=0.1 and b=0.3

*3) ETA Calculator*

*a) Notations*
   eta - running variable for ETA
   ETA(S,R)    - ETA for stop S and route id R

*b) Algorithm*
```
    for r ∈ Routes in database
    eta ← 0
    for l ∈ links in route r
    B ← set of buses in route r and link l
      if |B| != 0
      eta ← min(estimated time to link end for buses in B)
      else
          eta ← eta + TT_l
      if end(l) is stop S
          ETA(S,R) ← eta
```

**ACKNOWLEDGEMENTS**